\documentstyle[12pt]{article}
\def\be{\begin{equation}}
\def\ee{\end{equation}}
\def\bea{\begin{eqnarray}}
\def\eea{\end{eqnarray}}
\def\L{\Lambda}

\title{Black hole horizon in superstring inspired\\
$D$-dimensional dilaton gravity}
\author{E. G. Vorontsova, G. S. Sharov}
\date{}
\begin{document}
\maketitle
\centerline{\small
Tver State University,
Sadovyj per., 35, 170002, Tver, Russia.}
\vspace{3mm}

\begin{abstract}
In $D$-dimensional dilaton gravitational model stationary Schwarzschild type
solutions (centrally symmetrical solutions in vacuum)
are obtained. They contain the two-parameter set of solutions: the first
parameter is proportional to a central mass and the second one describes
intensity of the dilaton field. If the latter parameter equals zero
we have pure Schwarzschild solution with the black hole and constant
dilaton field. But when the mentioned parameter is non-zero and
the dilaton is non-constant, the black hole horizon vanishes.
\end{abstract}

\bigskip
\noindent{\bf Introduction}
\medskip

The considered dilaton gravitational model is the consequence of
the string theory in the low-energy limit \cite{FradTseyt,GSW}.
Here we study the generalization of the Schwarzschild solution
(that is stationary centrally symmetrical
solution in vacuum) for $D$-dimensional dilaton gravity.

We start from the following action \cite{FradTseyt}:
\be
S=\int\sqrt{|g|}\,e^{-2\phi(x)}\left[R+4(\nabla\phi)^2+\L
\right]\,d^{n+1}x.
\label{S}\ee
Here $\phi(x)$ is the scalar dilaton field in $n+1$\,-\,dimensional
($D=n+1$) pseudoriemann manifold $M_{1,n}$ with coordinates $x^\mu$
and the metric tensor $g_{\mu\nu}$, $g={}$det\,$|g_{\mu\nu}|$,
spatial dimensionality equals $n=2,3,\dots$, (the $1+1$ dimensional case
was studied previously in details), $R$ is the
scalar curvature, $(\nabla\phi)^2=g^{\mu\nu}\partial_\mu\phi\partial_\nu\phi$,
$\L$ is the constant. In Sects.~1,2 we concentrate on the case $\L=0$.

We use here and below the string frame for the action (\ref{S})
keeping in mind that it may be conformally transformed
$g_{\mu\nu}\;\to\;e^{-r\phi}g_{\mu\nu}$, $r=4/(D-2)$ to the Einstein frame
\cite{FradTseyt,Tseyt92,GasperVenez93}.

The equations of evolution for this model
\be\begin{array}{c}
R_{\mu\nu}+2\phi_{;\mu\nu}=0, \\
R+4(\nabla\phi)^2=\L.\rule[3mm]{0mm}{1mm}\end{array}
\label{sys}\ee
are obtained from action (\ref{S}) by the variation
of the gravitational field $g_{\mu\nu}$ and the dilaton field $\phi$
accordingly. Here $R_{\mu\nu}$ is the Ricci tensor, $\phi_{;\mu\nu}$ is the
covariant derivative.

\bigskip
\noindent{\bf 1. Schwarzschild-type solutions, $\L=0$}
\medskip

We search stationary Schwarzschild-type solutions of Eqs.~(\ref{sys}),
that is centrally symmetrical solutions in vacuum.
So the metric in the pseudoriemannian spherically symmetric
manifold $M_{1,n}$ with signature $+,-,-,\dots$ may be taken in the form
\bea
&&ds^2=e^{2\alpha(r)}dt^2-e^{2\beta(r)}dr^2- \nonumber \\
&&-r^2\bigl[{d\theta_1}^2+{\sin^2}\theta_1{d\theta_2}^2+\dots
+{\sin^2}\theta_1\cdot{\sin^2}\theta_2\dots\cdot{\sin^2}\theta_{n-2}
{d\theta_{n-1}}^2\bigr]. \label{metr}
\eea
Here $t$, $r$ and $\theta_i$ are time-like, radial and angular coordinates
accordingly.

We suppose below $n\ge 2$. The case $n=1$ resulting at $\L>0$
in ``two-dimensional" black hole $ds^2=\tanh^2{r\sqrt{\L}}\,dt^2-dr^2$, was
investigated in detail previously \cite{CallanGHS}.

When we substitute this metric and the corresponding nonzero Ricci tensor components
\bea
&R_{00}=e^{2\alpha-2\beta}[\alpha''+(\alpha')^2-\alpha'\beta']+
(n-1)\alpha'/r,&\nonumber\\
&R_{11}=[-\alpha''-(\alpha')^2+\alpha'\beta']+(n-1)\beta'/r,& \nonumber \\
&R_{22}=e^{-2\beta}[\beta'r-\alpha'r-(n-2)]+(n-2),& \nonumber \\
&R_{33}= R_{22}\sin^2\theta_1,\,\dots\, R_{nn}=R_{22}\sin^2\theta_1\sin^2
\theta_2\dots\sin^2\theta_{n-2}.&\nonumber
\eea
into the system (\ref{sys}), it reduces to the following four nontrivial
equations
\be\begin{array}{c}
\alpha''+\alpha'^2-\alpha'\beta'+(n-1)\alpha'/r=2\alpha'\phi', \\
\alpha''+\alpha'^2-\alpha'\beta'-(n-1)\beta'/r=2\phi''-2\beta'\phi', \\
\beta'-\alpha'+(n-2)(e^{2\beta}-1)/r+2\phi'=0, \\
\alpha''+\alpha'^2-\alpha'\beta'+(n-1)[(\alpha'-\beta')r+\frac12(n-2)
(1- e^{2\beta})]/r^2=2\phi'^2+\frac12\L e^{2\beta}.
\end{array}
\label{eq1}
\ee

Excluding $\beta'$ from these equations we obtain in the case $\L=0$ the following two relations:
\bea
&
1+(n-2)e^{2\beta}=-r\alpha''/\alpha',&
\label{al}\\
& 1+(n-2)e^{2\beta}=-r\phi''/\phi'.&
\label{ph}
\eea
Here $\alpha'\ne0$ and $\phi'\ne0$ are supposed.

Comparing expressions (\ref{al}) with (\ref{ph}) we see, that
\be
\phi'=Q\alpha',
\label{sw}
\ee
here $Q$ is the arbitrary constant.

In the case $n=2$ Eqs.~(\ref{al}) and (\ref{ph}) can be easily solved.
This solution has the form
\be
 ds^2=\left(\frac{r}{r_1}\right)^{a}dt^2-\left(\frac{r}{r_2}\right)^{b}dr^2-
r^2d\theta^2,\quad
\phi=\frac{a-b}4\ln\frac r{r_3}.
\label{n2}
\ee
Here $a$, $b$, $r_1$, $r_2$, $r_3$ there are constants, $a$ and $b$
are connected in the following way:
$$a^2=b^2+4b.$$
For arbitrary values of these parameters in the case $n=2$ we
have no singularities of the black hole (BH) type.
The case $b=0$ corresponds to the known conic solution
\be
 ds^2=dt^2-\mbox{const}\,dr^2-r^2d\theta^2,\qquad
\phi=\phi_0.
\label{conic}
\ee
in the standard gravitational model without dilaton ($\phi={}$const).

For $n\ge 3$ the solution of system (\ref{eq1}) is found in parametrical
form with using the parameter $\gamma=r\alpha'$:
\be\begin{array}{c}
e^{2(\alpha-\alpha_0)}=\big|(\gamma-\gamma_+)\big/(\gamma-\gamma_-)\big|
^{1/K}, \qquad e^{2\beta}=\big|\frac{4Q(Q-1)}{n-1}(\gamma-\gamma_+)(\gamma-
\gamma_-)\big|,\\
\phi=Q\alpha(r)+\phi_0,\qquad
r=r_0{\vert\gamma-\gamma_+\vert}^{A_+}{\vert\gamma-\gamma_-\vert}^{A_-}
|\gamma|^{-1/(n-2)}. \end{array}
\label{solut}
\ee
Here
$K=\sqrt{1+\tilde Q}$, $\tilde Q=4Q(Q-1)\big/(n-1)$,
\be
 \gamma_{\pm}=\frac{2Q-1\mp K}{\tilde Q},\qquad
A_{\pm}=\frac1{2(n-2)}\left(1\pm\frac{2Q-1}K\right).
\label{gam}
\ee

\bigskip
\noindent{\bf 2. Investigation of solutions}
\medskip

The geometrical and physical sense of obtained solutions (\ref{solut})
 was studied, in particular, with the help of test particles
in this gravitational field. If these particles do not interact directly
with the dilaton field $\phi$, their trajectories are
time-like geodesic lines. For example, the equation of
radial ($d\theta_1=\dots=d\theta_{n-1}=0$) geodesic lines for this metric are
$$
\frac{d^2t}{ds^2}+2\alpha'\frac{dt}{ds}\frac{dr}{ds}=0,\quad
\frac{d^2r}{ds^2}+\alpha'e^{2\alpha-2\beta}\left(\frac{dt}{ds}\right)^2+
\beta'\left(\frac{dr}{ds}\right)^2=0,
$$
and the corresponding physical velocity of a radially moving test particle
is
\be
v=\sqrt{\left|\frac{g_{11}}{g_{00}}\right|} \frac{dr}{dt}=
\pm \sqrt{1-e^{2\alpha-2C_0}}.
\label{v}\ee
Here $C_0={}$const.

Eq.~(\ref{v}) demonstrates that solutions (\ref{solut}) with $\alpha'>0$ or
$\gamma>0$ (remind that $\gamma=r\alpha'$) describe gravitational attraction
while negative $\gamma$ results in repulsing.
Taking this consideration into account we can extract the ``physical domain"
in the $Q\gamma$ plane (Fig.~1). This domain is bounded by the lines
$\gamma=0$ (the condition $\gamma>0$) and $\gamma=\gamma_+(Q)$. The latter
line is the $\gamma>0$ branch of the curve (\ref{gam}).
The solutions (\ref{solut}) corresponding to this domain describe
gravitational attraction and they are asymptotically flat at $r\to\infty$.

In the limit $Q\to0$ the solution (\ref{solut}) tends to the expression
\be
ds^2=\Big[1-(r_0/r)^{n-2}\Big]\,dt^2-(n-2)\Big[1-(r_0/r)^{n-2}\Big]^{-1}\,dr^2
-r^2(d\theta_1^2+\dots).
\label{Schw}\ee
and $\phi={}$const, that is the well-known Schwarzschild solution (for $n=3$).
The same formula (\ref{Schw}) may be obtained if we directly substitute
$\phi=Q=0$ into Eqs.~(\ref{eq1}). In this case the dilaton vanishes
($\phi={}$const).

So we may conclude that the solution (\ref{solut}) is the generalization
of the Schwarzschild solution for $D$-dimensional dilaton gravity.
The Schwarzschild solution (\ref{Schw}) is one-parameter family
with the parameter $r_g=r_0$ (gravitational radius proportional to
the central mass).
But in the case of the expression (\ref{solut}) we have the two-parameter
family of solutions: the parameter $r_0$ is the analog of $r_g$
and the additional parameter $Q$ demonstrates involving the
dilaton field in accordance with the expression (\ref{sw}).

And  the most interesting fact is that in this family of solutions
only dilaton-free solutions (\ref{Schw}) describe black holes (BH).
That is only in the case $Q=0$, $\phi={}$const we have the BH
horizon. When $Q\ne0$ and dilaton is non-trivial this horizon disappears.

This result may be explained by analyzing the expressions (\ref{solut}).
We are to investigate the limit of small $r$. It corresponds in the
the ``physical domain" of the $Q\gamma$ plane (Fig.~1) to the limit
$\gamma\to\gamma_+$ if $Q>0$ or $\gamma\to+\infty$ if $Q\le0$.
In the first case in this limit we have $r\to0$ and $g_{00}=e^{2\alpha}\to0$
(point of singularity),
and in the second case $Q<0$ the component $g_{00}=e^{2\alpha}$ tends to
non-zero constant when $r\to0$. The expression $|g_{11}|=e^{2\alpha}$
behaves correspondingly in these cases.

\bigskip
\noindent{\bf 3. $2+1$\,-\,dimensional solutions}
\medskip

For $n=2$ or $D=3$ the solutions of the system (\ref{eq1})
for an arbitrary value of $\L$ (including $\L\ne0$) were obtained.
In this case we may express $\phi'=(\alpha'-\beta')/2$ from the 3-rd
equaton (\ref{eq1}), substitute it into the first equaton and find
\be
e^{2\alpha(r)}=(r/r_1)^a,
\label{2al}\ee
where $a$ and $r_1$ are arbitrary constants.
The expression for $\alpha$ let us reduce the remaining equatons (\ref{eq1})
to the system
$$
\beta''-\frac{\beta'}r-(\beta')^2+\frac{a^2}{4r^2}=0,\quad
\beta''+\frac{\beta'}r=-\L e^{2\beta(r)}
$$
with the following solution:
\be
e^{2\beta(r)}=\frac{-(a^2+4)Cr^{B_\pm-2}}{\L(r^{B_\pm}-C)^2},
\qquad\phi(r)=\phi_0-\frac a{2r^2}-\frac{\beta(r)}2.
\label{2be}\ee
Here $B_\pm=\pm\sqrt{a^2+4}$, $C\equiv r_0^{2B_\pm}$ is the constant
of integration.
This expression $e^{2\beta(r)}$ is positive (physical) of
$C>0$, $\Lambda<0$ or $C<0$, $\Lambda>0$.
The first case is equivalent to the $2+1$\,-\,dimensional solutions
obtained in Ref.~\cite{Cadoni}, but in the latter case the solutions
(\ref{2al}), (\ref{2be}) differ from ones in Ref.~\cite{Cadoni}
because they are defined in $0<r<\infty$. Unlike the expressions \cite{Cadoni}
the solutions (\ref{2al}), (\ref{2be}) with $C<0$, $\Lambda>0$
have no singularities and cylindrical asymptotic behavior at large $r$.

Naturally, the expressions (\ref{2al}), (\ref{2be}) transform into
solutions (\ref{n2}) in the limit $\L\to0$ (with $\L/C\to{}$const).

\bigskip
\noindent{\bf Conclusion}
\medskip

In this work the stationary centrally symmetrical solutions in vacuum
(Schwarzschild type solutions) are obtained in the dilaton gravitational model
(\ref{S}) for different dimensionalities $D=n+1$. They generalize
the Schwarzschild solution (\ref{Schw}) and the conic $2+1$\,-\,dimensional
solution (\ref{conic}). These generalizations contain expressions (\ref{Schw}),
(\ref{conic}) because the Einstein gravity is the particular case of
the dilaton model (\ref{S}) when the dilaton field $\phi={}$const.
But it is non-trivial fact that any presence of the dilaton field
(that is $\phi'\ne0$, $Q\ne0$) radically changes the geometry.
In particular, the BH horizon vanishes in any stationary centrally symmetrical
solution (\ref{solut}) with $\L=0$ and $\phi'\ne0$.

\bigskip

\noindent{\bf Acknowledgements.}
The work is supported by the Russian Foundation of Basic Research
(grant 00-02-17359).

\end{document}